\documentclass[aps,prc,
	superscriptaddress,
	showpacs,preprintnumbers,amsmath,amssymb,nofootinbib,
	longbibliography
	]{revtex4-2}
\PassOptionsToPackage{dvipsnames}{xcolor}

\usepackage[
 	colorlinks=true,
	pdfstartview=FitV,
	linkcolor=blue,
	citecolor=magenta,
	urlcolor=blue,
	bookmarks=true,
	bookmarksnumbered=true,
	pdftitle={},
	pdfauthor={}
]{hyperref}

\usepackage{bbm,bm,mathtools,color,slashed}
\usepackage{amssymb,amsmath}
\usepackage[T1]{fontenc}

\usepackage[all]{xy}

\usepackage{graphicx}

\usepackage[framemethod=tikz]{mdframed}
\usepackage{tikz}

\usepackage[normalem]{ulem}  

\usepackage[most]{tcolorbox}

\newcommand{\cout}[1]{ \if 0 {#1} \fi }

\def\m@thcombine#1#2{%
  \setbox0=\hbox{$#1$}
  \setbox1=\hbox{$#2$}
  \ifdim\wd0>\wd1
    \setbox0=\hbox to\wd1{\hss\box0\hss}
  \else
    \setbox1=\hbox to\wd0{\hss\box1\hss}
  \fi
  \mathop{\vcenter{
    \offinterlineskip\box0\box1}}}
\def\lesim{\m@thcombine<\sim}
\def\gesim{\m@thcombine>\sim}

\def\lessgtr{\m@thcombine<>}
\def\gtrless{\m@thcombine><}
\newcommand{\bra}[1]{\left\langle #1 \right|}
\newcommand{\ket}[1]{\left| #1 \right\rangle}

\newcommand{\braket}[3]{\left\langle #1 \left| #2 \right| #3 \right\rangle}
\newcommand{\braketV}[3]{\left\langle #1 \left| #2 \right| #3 \right\rangle_V}
\newcommand{\braketpV}[3]{\left\langle #1 \left| #2 \right| #3 \right\rangle_{\mathrm{eff},V}}
\newcommand{\braketS}[3]{\left\langle #1 \left| #2 \right| #3 \right\rangle_{S}}
\newcommand{\braketpS}[3]{\left\langle #1 \left| #2 \right| #3 \right\rangle_{\mathrm{eff},S}}

\newcommand{\rbraketS}[3]{\left\langle #1 || #2 || #3 \right\rangle_{S}}
\newcommand{\rbraketpS}[3]{\left\langle #1 || #2 || #3 \right\rangle_{\mathrm{eff},S}}

\newcommand{\vecr}{\bm{r}}

\newcommand{\vecR}{\bm{R}}

\newcommand{\vecQ}{\bm{Q}}

\newcommand{\veck}{\bm{k}}
\newcommand{\vecK}{\bm{K}}
\newcommand{\vecG}{\bm{G}}

\newcommand{\vecxi}{\bm{\xi}}

\newcommand{\del}{\partial}

\newcommand{\beq}{\begin{equation}}
\newcommand{\eeq}{\end{equation}}
\newcommand{\beqa}{\begin{eqnarray}}
\newcommand{\eeqa}{\end{eqnarray}}
\newcommand{\bal}{\begin{align}}
\newcommand{\eal}{\end{align}}

\begin{document}

\title{
Interaction between nuclear clusters and superfluid phonons \\
in the neutron-star inner crust
}

\author{Masayuki Matsuo}
\email{matsuo@phys.sc.niigata-u.ac.jp}
\affiliation{Department of Physics, Faculty of Science, Niigata University, Niigata 950-2181, Japan}
\affiliation{Research Center for Nuclear Physics, Osaka University, Ibaraki  567-0047, Japan}

\author{Arata Nishiwaki} 
\author{Toshiyuki Okihashi}
\affiliation{Graduate School of Science and Technology, Niigata University, Niigata 950-2181, Japan}

\author{Masaru Hongo}
\email{hongo@phys.sc.niigata-u.ac.jp}
\affiliation{Department of Physics, Faculty of Science, Niigata University, Niigata 950-2181, Japan}
\affiliation{RIKEN Center for Interdisciplinary Theoretical and Mathematical Sciences (iTHEMS), RIKEN, Wako 351-0198, Japan}

\date{\today}

\begin{abstract}
The interaction between lattice vibrations of nuclear clusters and superfluid phonons associated with neutron superfluidity plays an important role in the dynamics of the neutron-star inner crust.
While this coupling has been discussed mainly within macroscopic approaches such as hydrodynamics and effective field theory, its microscopic origin and the value of the effective coupling constant have remained unclear.
In this work, we derive the interaction between nuclear clusters and superfluid phonons starting from a microscopic description of inner-crust matter.
Using nuclear density functional theory, we analyze the response of a neutron superfluid around a single nuclear cluster within the quasiparticle random-phase approximation.
From this microscopic response, we obtain the interaction between the cluster and the surrounding superfluid.
Matching this result to the long-wavelength effective description, we determine the coupling constant in an effective Hamiltonian describing the mixing between lattice and superfluid phonons.
The resulting coupling strength is found to be significantly smaller than previous hydrodynamical estimates.
This reduction originates from the suppression of the superfluid phonon amplitude inside and around the nuclear cluster.
Our results provide a microscopic determination of the coupling parameter governing lattice-superfluid phonon mixing in the neutron-star inner crust.
\end{abstract}
\maketitle

\section{Introduction}

The inner crust of neutron stars is a complex quantum many-body system, which consists of a lattice of nuclear clusters immersed in a neutron liquid and coexisting with relativistic degenerate electrons~\cite{Chamel:2008ca,Haensel:2007yy,Pethick:1995di}.
The neutrons are expected to exhibit superfluidity, which might play a key role in various astronomical phenomena~\cite{Chamel:2008ca,Dean:2002zx,Sedrakian:2018ydt}.
The quantum vortices emerging in the neutron superfluid carry a part of the total angular momentum of rotating stars, and their pinning and unpinning to the lattice has been argued to cause the glitch, a sudden change in the rotational frequency of pulsars~\cite{Anderson-Itoh1975,Alpar1977,Pines-Alpar92}.

Recently, the superfluid phonon, i.e., the phonon excitation in the neutron superfluid, has attracted attention as this low-energy degree of freedom contributes to the heat capacity~\cite{Khan:2004it,Martin:2014jja}. 
In addition, it has been suggested to affect the thermal conductivity and the elastic properties of the inner crust through coupling with lattice phonons~\cite{Aguilera:2008ed,Pethick:2010zf,Cirigliano:2011tj,PageReddy2012,Chamel:2012ix,Chamel:2013cva}. 
The interplay between superfluid and lattice degrees of freedom may even lead to exotic crystalline structures~\cite{Kobyakov:2013eta,Kobyakov:2013pza}. 
It is of critical importance to clarify the dynamics of superfluid phonons, especially their coupling to the lattice of nuclear clusters.

Previous studies of superfluid phonons have employed, on the one hand, macroscopic approaches such as hydrodynamic models~\cite{Pethick:2010zf,Kobyakov:2013eta,Kobyakov:2013pza,Chamel:2012ix,Chamel:2013cva,DiGallo:2011cr,Urban:2015dna} and low-energy effective field theory~\cite{Aguilera:2008ed,Cirigliano:2011tj}. 
These approaches are appropriate for describing the low-energy and long-wavelength limit of phonon dynamics~\cite{Aguilera:2008ed,Pethick:2010zf,Cirigliano:2011tj,PageReddy2012,Kobyakov:2013eta,Kobyakov:2013pza,Chamel:2012ix,Chamel:2013cva}. 
The parameters of these models are typically treated phenomenologically and determined from macroscopic quantities such as bulk thermodynamic properties.

On the other hand, microscopic approaches based on the many-body quantum theory of interacting nucleons have been developed within the framework of nuclear density functional theory. 
In particular, the linear response theory and the quasiparticle random phase approximation (QRPA) based on nuclear density functional theory or the Hartree--Fock--Bogoliubov (HFB) framework have been widely employed~\cite{Khan:2004it,Grasso:2008zz,Martin:2014jja,Inakura:2017fmp,Inakura:2018lim}.
These approaches have been successfully used to describe collective excitations of finite nuclei~\cite{RingSchuck1980,Paar:2007bk,Nakatsukasa:2016nyc}. 
An advantage of microscopic approaches is that they can describe nuclear clusters and the neutron superfluid in a unified way with common nucleon degrees of freedom, thus allowing one to explore the interaction between superfluid phonons and nuclear structure in detail.
For instance, in our previous studies of the dipole coupling of the superfluid phonon to nuclear clusters, we found that the superfluid phonon does not penetrate into the interior of nuclear clusters, suggesting that the coupling between superfluid and lattice phonons could be weak~\cite{Inakura:2017fmp,Inakura:2018lim}. 
It should be noted, however, that these studies adopt the Wigner--Seitz approximation, describing the system by a single Wigner--Seitz cell.

In the present paper, we extend our previous study to quantitatively evaluate the coupling between lattice and superfluid phonons in the inner crust.
To achieve this, we first formulate a general framework for linear-response density functional theory to describe superfluid phonons in inner-crust matter with a periodic lattice structure, and provide a scheme to construct an effective Hamiltonian describing the long-wavelength limit of lattice and superfluid phonons together with a phonon-phonon coupling represented by a single constant. 
However, a linear response calculation for inner-crust matter with a periodic lattice structure is a difficult task and, to the best of our knowledge, has not yet been achieved.  
We therefore propose a procedure in which the linear response method for a single-cell configuration is utilized in place of that for the periodic lattice configuration.  
This approximation may be justified provided that the interaction between a nuclear cluster and the superfluid phonon is local, i.e., mediated through a neutron–cluster potential well localized around the cluster. 
We also note that effects specific to the periodic lattice structure, such as the possible entrainment effect~\cite{Pethick:2010zf,Cirigliano:2011tj,PageReddy2012,Chamel:2012ix,Chamel:2013cva}, are not described.
Leaving such effects for future work, we treat in the present study the phonon-phonon coupling originating from the local cluster-phonon interaction.

The coupling constant obtained in the present work is significantly smaller than that obtained in macroscopic approaches.
This difference originates from the strong distortion of the phonon amplitude around the nuclear cluster obtained in the microscopic linear response density functional model.

The paper is organized as follows.
In Sec.~\ref{sec:formulation}, we introduce the theoretical framework to derive the effective phonon-phonon Hamiltonian. 
In Sec.~\ref{sec:calculation}, we explain a procedure to evaluate the phonon-phonon coupling constant using the linear response description of the superfluid phonon in a single-cell configuration, and present its numerical evaluation. 
In Sec.~\ref{sec:discussion}, we discuss the relation to previous studies based on effective field theory and hydrodynamical models, and quantitatively compare the phonon-phonon coupling constant.

\section{Formulation} 
\label{sec:formulation}

\subsection{Microscopic derivation of the effective coupled-phonon Hamiltonian}
\label{sec:microscopic-derivation}

It has been predicted that, in the shallow layers of the inner crust, 
a body-centered cubic (bcc) lattice of spherical nuclear clusters 
is immersed in a dilute neutron superfluid with an electron background.
We describe this system within the framework of nuclear density functional theory, in which the superfluidity (pairing correlation) is treated by the Bogoliubov--de Gennes equation.
In the traditional context of nuclear physics, this approach is also referred to as the Hartree--Fock--Bogoliubov (HFB) framework~\cite{Nakatsukasa:2016nyc}.

We consider a lattice of nuclear clusters in the inner crust and denote the center positions of the clusters by $\{ \vecR_i \}$, where the index $i$ runs over the clusters. 
The equilibrium positions are denoted by $\vecR_i^{(0)}$. 
When lattice phonons are excited, the cluster positions are displaced from the equilibrium positions as
\begin{equation}
 \hat{\vecR}_i(t) = \vecR_i^{(0)} + \hat{\vecQ}_i(t).
\end{equation}
For lattice phonons with wave vector $\veck$, polarization index $\lambda$, and frequency $\omega_{\veck\lambda}$, the displacement operator $\hat{\vecQ}_i(t)$ can be expanded in phonon modes as
\begin{equation}
\hat{\vecQ}_i(t)
= i \sum_{\veck,\lambda}
\left(\frac{\hbar}{2N_{\mathrm{cl}}M\omega_{\veck\lambda}}\right)^{1/2}
\vecxi_{\veck\lambda}
\left(
\hat{a}_{\veck\lambda} e^{-i\omega_{\veck\lambda} t}
+
\hat{a}_{-\veck\lambda}^{\dag} e^{i\omega_{\veck\lambda} t}
\right)
e^{i\veck \cdot \vecR_i^{(0)}},
\label{Eq:lattice coordinate}
\end{equation}
where $M$ is the mass of the nuclear cluster and $N_{\mathrm{cl}}$ is the number of clusters in a volume $V$.
Here, $\vecxi_{\veck\lambda}$ denotes the polarization vector of the lattice phonon.
As usual, $\hat{a}_{\veck\lambda}$ and $\hat{a}_{-\veck\lambda}^{\dag}$ are annihilation and creation operators for the lattice phonon~\cite{Mahan1990}.

The nucleons in the inner crust experience a self-consistent mean field that depends on the positions of the nuclear clusters.
We denote the corresponding potential by $\hat{v}_{0}(\vecr,\{\vecR_i\})$.
When the clusters oscillate according to Eq.~(\ref{Eq:lattice coordinate}), the self-consistent potential acquires a time-dependent fluctuation.
To linear order in the displacement $\hat{\vecQ}_i(t)$, the induced single-particle potential acting on the nucleons is written as
\begin{equation}
 \hat{V}_{c}(t)
 \equiv
 \sum_i
 \hat{\vecQ}_i(t) \cdot
 \frac{\partial \hat{v}_0(\vecr, \{\vecR_i\})}{\partial \vecR_i}.
\label{Eq:coupling term}
\end{equation}
The potential $\hat{V}_c(t)$ generated by the lattice vibration will be shown to produce a coupling between lattice phonons and superfluid phonons.

To investigate $\hat{V}_{c}(t)$, we focus on the low-energy collective excitations of the neutron superfluid in the inner crust, namely the superfluid phonons.
The superfluid phonon in the crystalline inner crust is characterized by the crystal momentum (wave vector) $\vecK$ and the excitation energy $\hbar \omega_{\vecK}^{s}$.
When the system is considered in a finite volume $V$, the allowed values of $\vecK$ are discretized.
Introducing the creation and annihilation operators of the superfluid phonon, $\hat{O}^\dagger_{\vecK}$ and $\hat{O}_{\vecK}$, the single-superfluid-phonon state is written as
\begin{equation}
 \ket{\vecK}_V=\hat{O}^\dagger_{\vecK}\ket{0},
\end{equation}
where $\ket{0}$ denotes the vacuum satisfying $\hat{O}_{\vecK} \ket{0} = 0$.

We now express the operator appearing in Eq.~\eqref{Eq:coupling term}
in terms of the superfluid phonon operators.
Restricting ourselves to a subspace of the superfluid phonons, the operator in Eq.~\eqref{Eq:coupling term} can be expanded as
\begin{equation}
 \frac{\partial \hat{v}_0(\vecr, \{\vecR_i\})}{\partial \vecR_i}
 \simeq \sum_{\vecK}
 \left(
 \braket{\vecK}
 {\frac{\partial \hat{v}_0(\vecr,\{\vecR_i\})}{\partial \vecR_i}}
 {0}
 \hat{O}_{\vecK}^{\dagger}
 +
 \braket{0}
 {\frac{\partial \hat{v}_0(\vecr,\{\vecR_i\})}{\partial \vecR_i}}
 {\vecK}
 \hat{O}_{\vecK}
 \right),
\end{equation}
where we inserted the truncated completeness relation.
Note that restricting ourselves to the single-phonon sector allows us to approximate $\ket{\vecK}\bra{0} \simeq \hat{O}_{\vecK}^{\dagger}$ and $\ket{0}\bra{\vecK} \simeq \hat{O}_{\vecK}$.

The matrix element $\bra{0} \frac{\partial \hat{v}_0}{\partial \vecR_i} \ket{\vecK}_{V}$ can be simplified using Bloch's theorem.
Since the superfluid phonon state with crystal momentum $\vecK$ acquires a phase factor under lattice translation, we obtain
\begin{equation}
\braket{0}{\frac{\partial \hat{v}_0}{\partial \vecR_i}}{\vecK}_{V} 
= e^{i\vecK\cdot\vecR_i^{(0)}}
\braket{0}{\frac{\partial \hat{v}_0}{\partial \vecR_0}}{\vecK}_{V} 
= - e^{i\vecK\cdot\vecR_i^{(0)}}
\braket{0}{\left.\frac{\partial \hat{v}_0}{\partial \vecr}\right|_{\vecR_0}}{\vecK}_{V},
\end{equation}
where $\vecR_0^{(0)}=0$ denotes the position of the reference cluster at the origin.
In the second equality, we used that the potential depends on the relative coordinate with respect to the cluster position.
Thus, the matrix element for an arbitrary cluster $i$ is reduced to the matrix element of the potential gradient $\left.\frac{\partial \hat{v}_0}{\partial \vecr}\right|_{\vecR_0}$ evaluated for the reference cluster.

Substituting the above relation into the coupling operator and carrying out the sum over the lattice sites, we obtain
\begin{align}
\hat{V}_c
&=
-i \sum_{\veck\lambda\vecG}
\left(\frac{N_{\mathrm{cl}}\hbar}{2M\omega_{\veck\lambda}}\right)^{1/2}
\left(\hat{a}_{\veck\lambda}
+ \hat{a}_{-\veck\lambda}^{\dagger}\right)
\notag \\
&\quad \times
\vecxi_{\veck\lambda} \cdot
\left[
\braketV{\veck+\vecG}{\left.\frac{\partial \hat{v}_0}{\partial \vecr}\right|_{\vecR_0^{(0)}}}{0}
\hat{O}^{\dagger}_{\veck+\vecG}
+
\braketV{0}{\left.\frac{\partial \hat{v}_0}{\partial \vecr}\right|_{\vecR_0^{(0)}}}{-\veck-\vecG}
\hat{O}_{-\veck-\vecG}
\right]
\notag  \\
 &\simeq 
 -i \sum_{\veck\lambda}
 \left(\frac{N_{\mathrm{cl}}\hbar}{2M\omega_{\veck\lambda}}\right)^{1/2}
 \left(\hat{a}_{\veck\lambda}
 + \hat{a}_{-\veck\lambda}^{\dagger}\right)
 \vecxi_{\veck\lambda} \cdot
 \left[
 \braketV{\veck}{\left.\frac{\partial \hat{v}_0}{\partial \vecr}\right|_{\vecR_0^{(0)}}}{0}
 \hat{O}^{\dagger}_{\veck}
 +
 \braketV{0}{\left.\frac{\partial \hat{v}_0}{\partial \vecr}\right|_{\vecR_0^{(0)}}}{-\veck}
 \hat{O}_{-\veck}
 \right],
\label{Eq:coupling operator}
\end{align}
where $\vecG$ denotes a reciprocal lattice vector resulting from the lattice sum.
In the last line, we neglected the umklapp processes with $\vecG\neq0$.
Because the coupling appears through the scalar product with the polarization vector $\vecxi_{\veck\lambda}$, the transverse lattice phonons do not couple to the superfluid phonon.

Combining the coupling operator derived above with the free Hamiltonians of the lattice phonon and the superfluid phonon, we obtain the Hamiltonian describing the coupled system:
\begin{align}
\hat{H} 
&= \sum_{\veck} \hbar\omega_{\veck}^{l}
\left( \hat{a}_{\veck}^{\dagger}\hat{a}_{\veck}+\frac{1}{2} \right) 
+
\sum_{\veck} \hbar\omega_{\veck}^{s}
\left( \hat{O}_{\veck}^{\dagger}\hat{O}_{\veck}+\frac{1}{2} \right) 
\notag \\
&\quad -
i \sum_{\veck}
\left(\frac{N_{\mathrm{cl}}\hbar}{2M\omega_{\veck}}\right)^{1/2}
\left(\hat{a}_{\veck}
+ \hat{a}_{-\veck}^{\dagger}\right)\frac{\veck}{|\veck|}
\left( 
\braketV{\veck}{\left.\frac{\del \hat{v}_0}{\del \vecr}\right|_{\vecR_0^{(0)}}}{0} \hat{O}^{\dagger}_{\veck} 
+ \braketV{0}{\left.\frac{\del \hat{v}_0}{\del \vecr}\right|_{\vecR_0^{(0)}}}{-\veck} \hat{O}_{-\veck}
\right),
\label{Eq:phonon Hamiltonian}
\end{align}
where the transverse lattice phonons are absent since they do not couple to the superfluid phonon, and we used $ \vecxi_{\veck l} = \veck/|\veck|$.
The Hamiltonian in Eq.~(\ref{Eq:phonon Hamiltonian}) is written solely in terms of the phonon operators, while the microscopic structure is encoded in the matrix elements $\braketV{\veck}{\left.\frac{\del \hat{v}_0}{\del \vecr}\right|_{\vecR_0^{(0)}}}{0}$ and $\braketV{0}{\left.\frac{\del \hat{v}_0}{\del \vecr}\right|_{\vecR_0^{(0)}}}{-\veck}$.

\subsection{Long-wavelength limit of the coupled-phonon Hamiltonian}
\label{sec:long-wave-length-limit}

We now construct an effective Hamiltonian describing the coupling between the lattice phonon and the superfluid phonon in the long-wavelength limit ($k \rightarrow 0$).
The microscopic phonon modes derived in the previous subsection are replaced by effective phonon degrees of freedom appropriate for this regime.

We introduce the creation and annihilation operators $\hat{b}_{\veck}^{\dagger}$ and $\hat{b}_{\veck}$ for the effective superfluid phonon, and define the corresponding one-phonon state in a volume $V$ by
\begin{equation}
 \ket{\veck}_{\mathrm{eff},V} = \hat{b}_{\veck}^{\dagger}\ket{0}.
 \label{eq:single-phonon-EFT}
\end{equation}
The matrix element of the density operator between the ground state and the one-phonon state is then given by (see Appendix~\ref{sec:note})
\begin{equation}
 \delta\rho_{\veck}^{\mathrm{eff},V}(\vecr) 
 \equiv
 \braketpV{0}{\hat{\rho}(\vecr)}{\veck}
 =
 -i \frac{1}{V^{1/2}}
 \sqrt{\frac{\hbar\omega_{\veck}^{s}}
 {2\frac{\del\mu_n}{\del \rho_n}}}
 e^{i\veck\cdot\vecr}.
\end{equation}
Here $\mu_n$ denotes the neutron chemical potential, and
\begin{equation}
\frac{\del\mu_n}{\del \rho_n}
=
\frac{\del^2 E_n}{\del\rho_n^2}
\end{equation}
is the second derivative of the neutron energy density $E_n$ with respect to the neutron density $\rho_n$.
This quantity is related to the compressibility of the neutron superfluid, and the superfluid-phonon velocity is accordingly given by
\begin{equation}
 v_s = \sqrt{\frac{\rho_n }{m_n} \frac{\del\mu_n}{\del \rho_n}},
 \label{eq:vs-superfluid-phonon}
\end{equation}
where $m_n$ is the neutron mass.

In the microscopic Hamiltonian derived in the previous subsection,
the operators $\hat{O}_{\veck}^{\dagger}$ and $\hat{O}_{\veck}$ describe phonon excitations in the presence of the nuclear cluster.
The corresponding phonon modes are therefore not simple plane waves.
In the long-wavelength limit, we approximate these microscopic phonon operators by the plane-wave phonon operators $\hat{b}_{\veck}^{\dagger}$ and $\hat{b}_{\veck}$ introduced above.
This replacement amounts to projecting the microscopic phonon excitations onto the plane-wave phonon states of the uniform neutron superfluid.

The interaction between the nuclear cluster and the neutrons must therefore be reformulated in the effective description.
In the microscopic Hamiltonian this interaction is described by the potential $\hat{v}_0$, which couples to the neutron density operator.
We therefore introduce an effective interaction
\begin{equation}
\hat{v}_{\mathrm{eff}}
=
\int v_{\mathrm{eff}}(\vecr)\,\hat{\rho}(\vecr)\, d\vecr ,
\end{equation}
which represents the coupling between the nuclear cluster and the neutron density in the effective theory.
Here we assume a local potential acting on neutrons.

The effective interaction $\hat{v}_{\mathrm{eff}}$ is defined so that the coupling matrix element between the ground state and the one-phonon state is reproduced in the effective description.
This requirement leads to the matching condition
\begin{equation}
\braketV{0}{\left.
\frac{\del \hat{v}_0}{\del \vecr}
\right|_{\vecR_0^{(0)}}}{\veck}
=
\braketpV{0}{\frac{\del \hat{v}_{\mathrm{eff}}}{\del \vecr}}{\veck},
\label{Eq:matching condition}
\end{equation}
where $\ket{\veck}_{\mathrm{eff},V}$ denotes the plane-wave superfluid phonon state with momentum $\veck$ in the uniform neutron superfluid without nuclear clusters.
In this formulation the modification of the phonon mode due to the presence of the nuclear cluster is encoded in the effective potential $\hat{v}_{\mathrm{eff}}$ rather than in the phonon state.
Using the density matrix element introduced above, the right-hand side of Eq.~(\ref{Eq:matching condition}) becomes
\begin{equation}
\braketpV{0}{\frac{\del \hat{v}_{\mathrm{eff}}}{\del \vecr}}{\veck}
=
\int
\frac{\del v_{\mathrm{eff}}(\vecr)}{\del \vecr}\,
\delta\rho_{\veck}^{\mathrm{eff},V}(\vecr)\,
d\vecr .
\end{equation}

Substituting the density fluctuation associated with the superfluid phonon into the effective interaction $\hat v_{\mathrm{eff}}$, we obtain the coupling between the lattice phonon and the superfluid phonon.
The interaction Hamiltonian between the lattice phonon and the superfluid phonon is then written in the effective description as
\begin{equation}
\hat{V}_{\mathrm{eff},c}=
\frac{1}{2}\sum_{\veck} \hbar g_{\veck}  
\sqrt{\frac{\hbar\omega_{\veck}^s}{\hbar\omega_{\veck}^l}}
\left( \hat{a}_{\veck}+\hat{a}_{-\veck}^{\dagger}\right)
i \left( \hat{b}_{\veck}^{\dagger}-\hat{b}_{-\veck}\right)
\label{Eq:interaction}
\end{equation}
where the momentum-dependent coupling constant is given by
\begin{equation}
g_{\veck}= 
\left( 
\frac{\rho_{\mathrm{cl}}}{ M\frac{\del\mu_n}{\del \rho_n}} 
\right)^{1/2} 
\bar{v}_{\mathrm{eff}}(\veck),
\quad \mathrm{with} \quad 
\bar{v}_{\mathrm{eff}}(\veck)=
\int e^{-i\veck\cdot \vecr} v_{\mathrm{eff}}(\vecr) d\vecr .
\end{equation}
Here, $\rho_{\mathrm{cl}}=N_{\mathrm{cl}}/V$ denotes the number density of nuclear clusters.

In the long-wavelength limit, the Fourier component $\bar{v}_{\mathrm{eff}}(\veck)$ can be expanded around $\veck=0$.
Keeping the leading contribution, we obtain an effective phonon--phonon Hamiltonian valid in the long-wavelength limit,
\begin{align}
\hat{H}_{\mathrm{eff}} 
&= \sum_{\veck} \hbar\omega_{\veck}^{l}
\left( \hat{a}_{\veck}^{\dagger}\hat{a}_{\veck}+\frac{1}{2} \right) 
+
\sum_{\veck} \hbar\omega_{\veck}^{s}
\left( \hat{b}_{\veck}^{\dagger}\hat{b}_{\veck}+\frac{1}{2} \right) 
\notag \\
&\quad +
\frac{1}{2}\sum_{\veck} \hbar g_0 k 
\sqrt{\frac{v_s}{v_l}}
\left( \hat{a}_{\veck}+\hat{a}_{-\veck}^{\dagger}\right)
i \left( \hat{b}_{\veck}^{\dagger}-\hat{b}_{-\veck}\right),
\label{Eq:Heff}
\end{align}
where the bare dispersion relations of the lattice and superfluid phonons and the coupling constant are given by
\begin{equation}
 \omega_{\veck}^{l}= v_{l} k, \quad
 \omega_{\veck}^{s}= v_{s} k, \quad
 g_0 = 
 \left( 
 \frac{\rho_{\mathrm{cl}}}{ M\frac{\del\mu_n}{\del \rho_n}} 
 \right)^{1/2} 
 \bar{v}_{\mathrm{eff}}(0).
\label{Eq:coupling constant}
\end{equation}
Here $v_l$ and $v_s$ denote the velocities of the lattice and superfluid phonons, respectively.
The interaction between the lattice and superfluid phonons is therefore characterized by the single constant $g_0$.
It is worth noting that the coupling term proportional to $g_0$ mixes the lattice and superfluid phonons, leading to hybridized collective modes.
The quantity 
\begin{equation}
 \bar{v}_{\mathrm{eff}}(0)=\int v_{\mathrm{eff}}(\vecr)d\vecr,
\end{equation}
is the volume integral of the effective interaction potential.

The remaining task is to determine the effective potential $\hat{v}_{\mathrm{eff}}$ so that the matching condition Eq.~(\ref{Eq:matching condition}) is satisfied.
This problem will be addressed in the next section.

For reference, it is instructive to consider a simple approximation in which the effective interaction is replaced by the original microscopic potential $\hat{v}_0$.
In this crude approximation the matching condition Eq.~(\ref{Eq:matching condition}) is not satisfied, because the microscopic phonon mode in the presence of the nuclear cluster differs from the plane-wave phonon mode used in the effective description.
In this approximation the coupling constant $\bar{v}_{\mathrm{eff}}(0)$ in
Eq.~(\ref{Eq:coupling constant}) would be replaced by
\begin{equation}
\bar{v}_0(0)
=
\int
\left(
\left.
v_0(\vecr)
\right|_{\vecR_0^{(0)}}
-
v_{0,\mathrm{uni}}
\right)
d\vecr ,
\end{equation}
which represents the volume integral of the neutron potential around the nuclear cluster located at $\vecR_0^{(0)}$, measured relative to the constant neutron potential $v_{0,\mathrm{uni}}$ of the uniform neutron superfluid.

\subsection{Phonon-phonon interaction in density functional theory}
\label{sec:interaction-DFT}

In this subsection, we discuss how the phonon-phonon interaction can be determined within the framework of density functional theory.
After specifying a way to compute the matrix element in the original Hamiltonian~\eqref{Eq:phonon Hamiltonian}, we consider the simplified setup in Eq.~\eqref{Eq:matching condition}.

The potential $\hat{V}_c(t)$ generated by the lattice vibrations acts as an external field on the nucleons in the inner crust.
The response of the system to this perturbation can be described within the linear response formalism of the time-dependent Kohn--Sham--Bogoliubov--de Gennes theory.
In this framework, the time evolution of the quasiparticle states is governed by the self-consistent time-dependent single-particle Hamiltonian
\begin{equation}
 \hat{h}(t) = \hat{T} + \hat{v}_{\mathrm{sc}}[\left\{ \rho_\alpha(t) \right\}] + \hat{V}_{c}(t),
\end{equation}
where $\hat{T}$ is the single-particle kinetic energy operator.
Here $\hat{v}_{\mathrm{sc}}[\{\rho_\alpha(t)\}]$ denotes the time-dependent self-consistent field, which is a functional of a set of densities $\{\rho_\alpha(t)\}$.
Expanding the density fluctuations $\delta\rho_\alpha(t)$ to linear order in the perturbation, one obtains the linear response equation for $\delta \rho_\alpha(\omega)$ in the energy domain.
The eigenmodes of this equation describe the normal modes of excitation in the inner crust.
We omit the details of the formalism and refer to Refs.~\cite{Matsuo:2001wy,Matsuo:2002gu,Nakatsukasa:2016nyc}.

The key quantity provided by the linear response formalism is the transition density associated with the superfluid phonon,
\begin{equation}
 \delta\rho_{\vecK}^{V}(\vecr)
 =
 \braketV{0}{\hat{\rho}(\vecr)}{\vecK},
\end{equation}
where $\ket{\vecK}_{V}$ denotes the superfluid phonon state and
\begin{equation}
 \hat{\rho}(\vecr)
 =
 \sum_{\sigma}
 \psi^{\dagger}(\vecr\sigma)\psi(\vecr\sigma)
\end{equation}
is the nucleon density operator.
Here $\psi^{\dagger}(\vecr\sigma)$ and $\psi(\vecr\sigma)$ denote
nucleon field operators with spin index
$\sigma=\uparrow,\downarrow$.

With the creation and annihilation operators of the superfluid
phonon, $\hat{O}^\dagger_{\vecK}$ and $\hat{O}_{\vecK}$ as well as the single-superfluid phonon state $\ket{\vecK}_V=\hat{O}^\dagger_{\vecK}\ket{0}$, we can expand the density operator in terms of the phonon operators as
\begin{align}
 \hat{\rho}(\vecr)
 &\simeq
 \sum_{\vecK}
 \left(
 \braketV{\vecK}{\hat{\rho}(\vecr)}{0}\hat{O}^\dagger_{\vecK}
 +
 \braketV{0}{\hat{\rho}(\vecr)}{\vecK}\hat{O}_{\vecK}
 \right)
 \notag \\
 &=
 \sum_{\vecK}
 \left(
 \delta\rho_{\vecK}^{V}(\vecr)^* \hat{O}^\dagger_{\vecK}
 +
 \delta\rho_{\vecK}^{V}(\vecr) \hat{O}_{\vecK}
 \right),
 \label{Eq:density operator}
\end{align}
where we have retained only the one-phonon contributions and omitted
the constant ground-state density.
Correspondingly, the matrix element of the neutron--cluster
interaction appearing in the microscopic Hamiltonian can be written as
\begin{equation}
\braketV{0}{\left. \frac{\del \hat{v}_0}{\del \vecr}\right|_{ \vecR_0^{(0)}}}{\vecK}
= \int \left.\frac{\del v_0(\vecr)}{\del \vecr}\right|_{\vecR_0^{(0)}} \delta\rho_{\vecK}^{V}(\vecr) d\vecr,
\label{Eq:coupling matrix element1}
\end{equation}
where $\delta\rho_{\vecK}^{V}(\vecr)$ denotes the transition density of the superfluid phonon with crystal momentum $\vecK$.
Here we assumed that 
the microscopic self-consistent potential can be written as 
a one-body potential
$\hat{v}_0 = \int d\vecr\, v_0(\vecr)\hat{\rho}(\vecr)$,
neglecting the pair potential in the self-consistent field $\hat{v}_0$.
This approximation is justified because the pair potential is typically of order $1\,\mathrm{MeV}$, much smaller than the one-body potential of order $50\,\mathrm{MeV}$.
Equation (\ref{Eq:coupling matrix element1}) shows that the coupling between the lattice vibration and the superfluid phonon in Eq.~\eqref{Eq:phonon Hamiltonian} can, in principle, be obtained once the transition density $\delta\rho_{\vecK}^{V}(\vecr)$ is determined within the linear response density functional theory.

In practice, however, performing such a calculation for the crystalline inner crust is a highly nontrivial task, since a self-consistent linear response calculation for a periodic lattice configuration of nuclear clusters is not currently available.
Moreover, even if such a calculation were feasible, the resulting expression would not provide a transparent description of the long-wavelength limit relevant for the effective theory developed in the previous subsection.

In the following section, we therefore determine the effective coupling
within a simplified approximation based on Eq.~\eqref{Eq:matching condition}.

\section{Determination of the effective potential}
\label{sec:calculation}

We discuss the determination of the effective interaction between a nuclear cluster and the superfluid phonon by matching the response of the effective theory to that of the microscopic theory in the long-wavelength limit.
In Sec.~\ref{sec:matching-single-cluster} we formulate the matching condition in a simplified setup where a single nuclear cluster is embedded in a large spherical box.
In Sec.~\ref{sec:effective-side} we evaluate the effective-theory side of the matching condition.
In Sec.~\ref{sec:microscopic-side} we evaluate the microscopic side using linear-response density functional theory.
In Sec.~\ref{sec:results} we present the resulting estimate of the effective coupling and discuss its physical implications.

\subsection{Matching condition in the single-cluster approximation}
\label{sec:matching-single-cluster}

In this subsection, we construct a practical implementation of the matching condition given in Eq.~\eqref{Eq:matching condition}.
Since a fully microscopic calculation of the phonon response in the periodic lattice is not presently available, we introduce a simplified setup in which the matching condition can be evaluated explicitly.

In this setup, the system is approximated by a single nuclear cluster embedded in a large spherical box, and the matrix element appearing in the matching condition \eqref{Eq:matching condition} is evaluated in this configuration.
Within this formulation, the matching condition equates the matrix element of the microscopic coupling operator with that of the effective interaction evaluated for the same phonon state.

We first consider the microscopic matrix element, Eq.(\ref{Eq:coupling matrix element1}), on the left-hand side of Eq.~\eqref{Eq:matching condition}.
The key observation is that the dominant contribution to the integral in Eq.~\eqref{Eq:coupling matrix element1} arises from the region around the nuclear cluster.
Indeed, the derivative of the potential, $\del v_0(\vecr)/\del \vecr$, is localized near the cluster surface and becomes small outside the cluster region provided that the nuclear clusters are sufficiently well separated.
Therefore, the integral is mainly determined by the transition density in the vicinity of the cluster.

In previous studies~\cite{Inakura:2017fmp,Inakura:2018lim}, where the superfluid phonon was described within the spherical Wigner--Seitz approximation, the phonon was found to be significantly distorted by the presence of the nuclear cluster.
More precisely, the phonon wave hardly penetrates into the cluster, while in the neutron superfluid region outside the cluster the phonon amplitude exhibits a simple wave pattern consistent with a standing wave in the cell.

It is therefore reasonable to expect that the superfluid phonon in the periodic lattice exhibits a similar structure near each nuclear cluster. 
Motivated by this observation, we approximate the transition density in the lattice system of volume $V$ by that obtained in a setup consisting of a single nuclear cluster embedded in a large spherical box $S$ with radius $R$ for the present calculation. 
In the original lattice system, the spacing between nuclear clusters is characterized by the Wigner--Seitz radius $R_{\mathrm{WS}}$, while the cluster size is denoted by $R_{\mathrm{clust}}$. 
We assume $R \gtrsim R_{\mathrm{WS}} \gg R_{\mathrm{clust}}$ so that the box boundary does not affect the phonon structure near the cluster. 
In this setup the system becomes spherically symmetric, and the phonon states can be described in the partial-wave representation and labeled by the radial wave number $k_n$ and the angular momentum quantum numbers $LM$. 
The validity of this approximation relies on the separation of scales between the cluster size and the lattice spacing, $R_{\mathrm{WS}} \gg R_{\mathrm{clust}}$, and we focus on the long-wavelength regime with $k_n R_{\mathrm{clust}} \ll 1$.

Within this configuration, the matching condition Eq.~\eqref{Eq:matching condition} is replaced by 
\begin{equation}
\braketS{0}{ \frac{\del \hat{v}_0}{\del \vecr}}{k_{n} LM}
= \braketpS{0}{\frac{\del \hat{v}_{\mathrm{eff}}}{\del \vecr}}{k_{n} LM}
 \quad \mathrm{for} \quad L=1
\label{Eq:matching condition2}
\end{equation}
where the left-hand side is evaluated using the linear-response density functional theory for a single cluster in the spherical box $S$, while the right-hand side is evaluated using the unperturbed superfluid phonon states together with the effective potential. 
The relevant partial wave is the dipole mode $L=1$ because $\del \hat{v}_0 / \del \vecr$ is a rank-one operator.

The determination of the effective potential therefore reduces to the evaluation of the two matrix elements appearing in Eq.~\eqref{Eq:matching condition2}. 
In the following subsections, we evaluate these matrix elements separately: the effective-theory side is obtained analytically using the phonon wave functions in the spherical box, while the microscopic side is computed numerically using linear-response calculations.

\subsection{Evaluation in the effective theory}
\label{sec:effective-side}

In this subsection, we evaluate the effective-theory side of the matching condition~\eqref{Eq:matching condition2}. 
The matrix element on the right-hand side of Eq.~\eqref{Eq:matching condition2} can be written, for the dipole channel $L=1$, in terms of the transition density of the phonon state as
\begin{equation}
 \braketS{0}{\frac{d \hat v_{\mathrm{eff}}}{dr} Y_{1M}}{k_n 1M}
 =
 \int d\vecr\,
 \frac{d v_{\mathrm{eff}}(r)}{dr}
 Y_{1M}(\hat{\vecr})
 \delta\rho_{k_n 1M}^{\mathrm{eff},S}(\vecr),
 \label{eq:matrix-element-EFT}
\end{equation}
where the transition density of the unperturbed superfluid phonon in the spherical box is given by (see Appendix~\ref{sec:note})
\begin{equation}
\delta\rho_{k_{n}LM}^{\mathrm{eff},S}(\vecr)  =  \frac{1}{R^{3/2}}\sqrt{\frac{\hbar\omega_{k_n}^{s}}{2\frac{\del\mu_n}{\del \rho_n}}}
c_{nL}j_L(k_nr)Y_{LM}(\hat{\vecr}).
\label{eq:trans_analytic}
\end{equation}
Here, $j_L$ denotes the spherical Bessel function and $Y_{LM}$ the spherical harmonics.

To proceed, we introduce a parametrized form of the effective potential.
For simplicity, we adopt a Gaussian form
\begin{equation}
 v_{\mathrm{eff}}(\vecr) = v_e e^{-r^2/b^2},
\end{equation}
which introduces two parameters: the strength $v_e$ and the range parameter $b$.
Since our interest lies in the low-momentum behavior, the detailed shape of the potential is not important, and the parameter $b$ may be chosen to be of the order of the nuclear-cluster size.

With these ingredients, the matrix element on the effective-theory side can be evaluated explicitly by performing the spatial integral.
The resulting expression is proportional to the strength parameter $v_e$, which will be determined by the matching condition once the microscopic matrix element is evaluated.

\subsection{Microscopic linear-response calculation}
\label{sec:microscopic-side}

We next evaluate the left-hand side of Eq.~\eqref{Eq:matching condition2} using the linear-response density functional theory (DFT) framework of Refs.~\cite{Inakura:2017fmp,Inakura:2018lim}, with several modifications described below.

\paragraph{DFT setup}
For the energy density functional, we adopt the Skyrme functional with the parameter set SLy4.
The pairing interaction is given by a density-dependent delta interaction
\begin{align}
v_{\rm pair,\tau}(\vecr,\vecr^{\prime})
 = v_{0} \frac{1-P_{\sigma}}{2} \left[ 1 - \eta \left( \frac{\rho_{\tau}(\vecr)}{\rho_{c}} \right)^{\alpha} \right] \delta(\vecr - \vecr^{\prime}),
\end{align}
where $P_{\sigma}$ denotes the spin-exchange operator and $\tau=n,p$ labels the nucleon isospin.
The density $\rho_\tau(\vecr)$ represents the local particle density of species $\tau$.
The parameters are taken as
$v_{0}=-458.4$ MeV,
$\rho_{c}=0.08$ fm$^{-3}$,
$\eta=0.845$, and
$\alpha=0.59$~\cite{Matsuo:2005vf,Matsuo:2010jj,Okihashi:2020kqg}.
These parameters are chosen to reproduce the neutron pairing gap in neutron matter obtained in BCS calculations with a bare nuclear force in the ${}^1S_0$ channel.

The quasiparticle space is truncated at a maximal quasiparticle energy
$E_{\rm max}=60$ MeV.
The box size is chosen as $R=50$ fm, which is sufficiently large compared with the size of the nuclear cluster.
All quasiparticle states within this space, covering partial waves up to $j=167/2$, are included in both the ground-state and linear-response calculations.

In the linear-response calculation we consider only neutron dynamics, while keeping the protons in their ground-state configuration, since the motion of the nuclear cluster is treated as an external perturbation.
The residual interaction is treated within the Landau--Migdal approximation.
The renormalization factor introduced in Refs.~\cite{Inakura:2017fmp,Inakura:2018lim} is not employed in the present study.

To handle the large box size required in the present calculation, we modified the numerical code of Refs.~\cite{Inakura:2017fmp,Inakura:2018lim}. 
In particular, the radial HFB equation is solved by a diagonalization method using a grid representation of the radial quasiparticle wave functions, following the implementation described in Ref.~\cite{Okihashi:2020kqg}.

\paragraph{Superfluid phonon mode in neutron superfluid}

In the following, we consider one representative setup, leaving a systematic analysis for future work.
We study a nuclear cluster containing $Z=28$ protons immersed in a neutron superfluid whose chemical potential is fixed to $\mu_n = 4$ MeV.
The ground-state configuration is the same as that used in Ref.~\cite{Okihashi:2020kqg}.
For comparison, we also perform the linear-response calculation for a configuration in which the nuclear cluster is absent.
This configuration corresponds to a uniform neutron superfluid described by the same energy density functional, whose ground state is obtained from BCS calculations for neutrons.
The resulting quantities for several representative values of $\mu_n$ are listed in Table~\ref{table-uniform BCS}.

\begin{figure}[t]
\centering
\includegraphics[width=0.4\columnwidth]{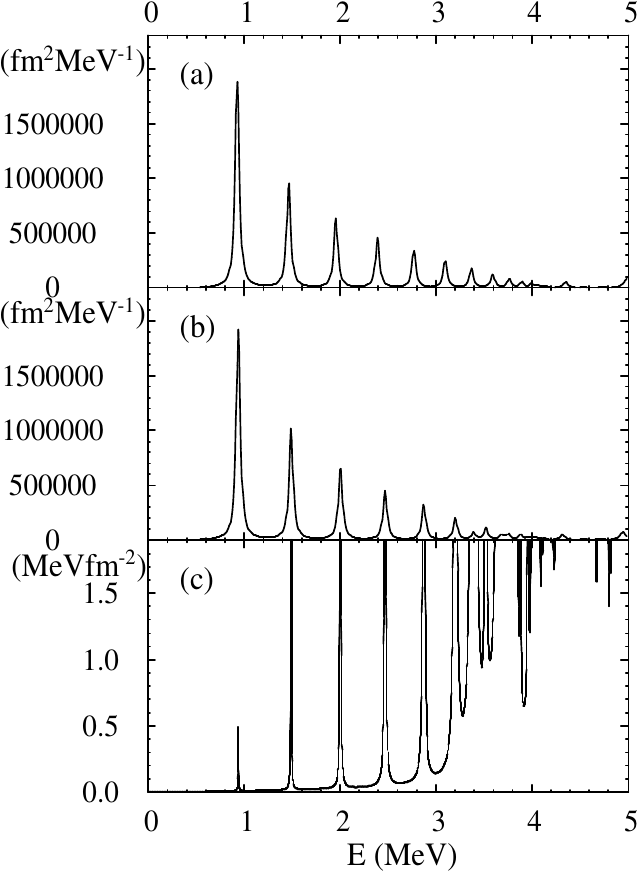}
 \caption{
  Strength functions in neutron superfluid matter with chemical potential $\mu_n=4\,\mathrm{MeV}$ 
  confined in a spherical box with radius $R=50\,\mathrm{fm}$, with and
  without the presence of a nuclear cluster.
  (a) Strength function of the dipole operator $rY_{1M}$  without a nuclear cluster.
  The smoothing width is $50\,\mathrm{keV}$.
  (b) Same as (a), but for the configuration with a nuclear cluster with proton number $Z=28$ embedded in the neutron superfluid matter. 
  (c) Same as (b), but for the coupling operator $\frac{dU_{\mathrm{Skyrme}}}{dr}Y_{1M}$ for the configuration with a nuclear cluster.
  Here 
  the smoothing width is $5\,\mathrm{keV}$.
  }
\label{fig-strength_function}
\end{figure}

\begin{figure}[t]
\centering
\includegraphics[width=0.4\columnwidth]{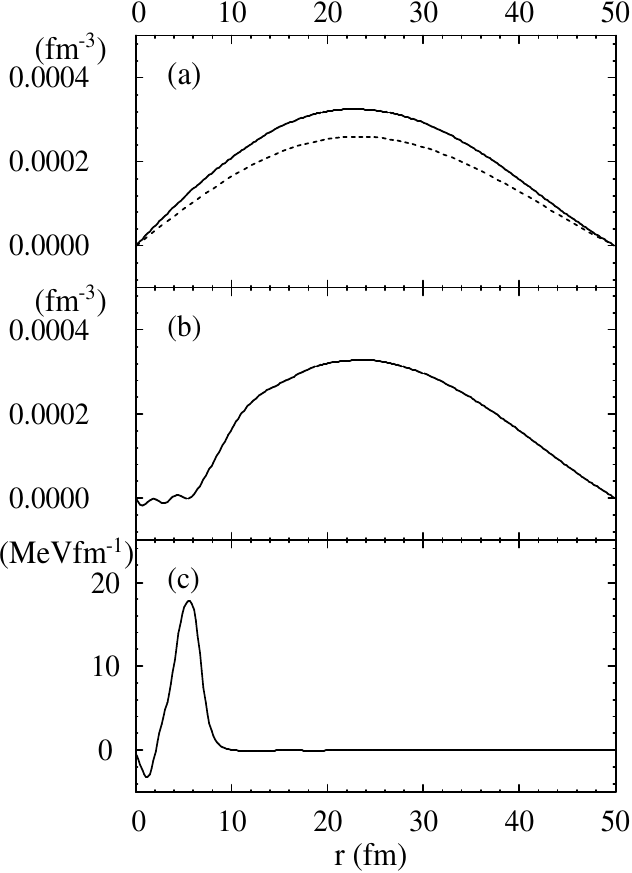}
\caption{
  (a) Transition density for the superfluid phonon state with the lowest excitation energy in a uniform neutron superfluid with $\mu_n=4$ MeV confined in a spherical box with $R=50$ fm.
  The solid curve corresponds to the peak at $E=0.931$ MeV in Fig.~\ref{fig-strength_function}\,(a).
  The dotted curve shows the analytic result obtained from Eq.~\eqref{eq:trans_analytic}.
  (b) Transition density for the superfluid phonon state with the lowest excitation energy for the configuration in which a nuclear cluster with proton number $Z=28$ is embedded in the neutron superfluid with $\mu_n=4$ MeV.
  The state corresponds to the peak at $E=0.941$ MeV in Fig.~\ref{fig-strength_function}\,(b).
  (c) Coupling potential $dU_{\mathrm{Skyrm}}/dr$ for the configuration with the nuclear cluster.
  }
\label{fig-transition_density}
\end{figure}

Figure~\ref{fig-strength_function}\,(a) shows the strength function for the dipole operator $rY_{1M}$,
\begin{equation}
S_r(\hbar\omega) = \sum_{n,M} 
\left| \braketS{n,L=1,M}{rY_{1M}}{0} \right|^2 \delta(\hbar\omega-\hbar\omega_{n,L=1})
=\sum_n \left| \rbraketS{n,L=1}{rY_1}{0} \right|^2 \delta(\hbar\omega-\hbar\omega_{n,L=1}),
\end{equation}
calculated for the configuration without a nuclear cluster.
Several low-lying peaks located below $\hbar\omega \lesssim 4~\mathrm{MeV} \approx 2\Delta$ correspond to superfluid phonon excitations.
Note that the dipole matrix element appearing in the strength function is related to the transition density as 
\begin{equation}
\delta\rho_{nLM}(\vecr)
=
\braketS{nLM}{\hat{\rho}(\vecr)}{0}
\end{equation}
through
\begin{equation}
\braketS{n,L=1,M}{rY_{1M}}{0}
=
\int d\vecr\,
r Y_{1M}(\hat{\vecr})
\,
\delta\rho_{n1M}(\vecr).
\end{equation}
For the lowest peak, the excitation energy, dipole matrix element, and transition density are summarized in Table~\ref{table-strength} and in Fig.~\ref{fig-transition_density}\,(a).
These quantities are compared with the corresponding results obtained from the analytic phonon solution given in Eq.~\eqref{eq:trans_analytic}.

We find that the microscopic linear-response results agree reasonably well with the analytic prediction, although small differences remain.
The transition density is nearly proportional to the analytic form $j_1(kr)$.
However, the amplitude of the transition density and the dipole strength exceed the analytic values by factors of 1.25 and 1.55, respectively.
The excitation energy differs by about $6\%$.
These differences may originate from the Landau--Migdal approximation adopted for the residual interaction or from the truncations introduced in the numerical calculation.

Figure~\ref{fig-strength_function}\,(b) shows the dipole strength function for the configuration in which a nuclear cluster with $Z=28$ is placed at the origin of the box.
The spectrum is very similar to that obtained without the nuclear cluster.
The excitation energies and dipole matrix elements of the superfluid phonon modes, listed in Table~\ref{table-strength}, are only weakly affected by the presence of the cluster.
This indicates that the presence of the cluster has little effect on the global dipole response.

However, as seen in Fig.~\ref{fig-transition_density}\,(b), the transition density inside and around the nuclear cluster ($r \lesssim 10\,\mathrm{fm}$) is significantly distorted, with a substantial reduction of its amplitude, as already observed in Ref.~\cite{Inakura:2017fmp}.
In particular, the transition density almost vanishes inside the nuclear cluster ($r \lesssim 6\,\mathrm{fm}$).
In contrast, far outside the cluster the transition density is essentially the same as that of a superfluid phonon not interacting with the cluster.
This behavior is consistent with the observation that the excitation energy and the dipole matrix element are almost identical to those obtained in the absence of the cluster.

\paragraph{Cluster-phonon coupling}
To determine the cluster-phonon coupling appearing in the matching condition Eq.~\eqref{Eq:matching condition2}, we calculate the strength function of the operator $\frac{dU_{\mathrm{Skyrm}}}{dr} Y_{1M}$, where $U_{\mathrm{Skyrm}}(\vecr)$ denotes the central part of the self-consistent Skyrme-Hartree-Fock potential for neutrons as $U_{\mathrm{Skyrm}}(\vecr)$
is a dominant part of the interaction $v_0(\vecr)$ between the nuclear-cluster and neutrons.
The corresponding strength function is defined as
\begin{align}
 S_{dU}(\hbar\omega) 
 &\equiv \sum_{n,M} \left| \braketS{n,L=1,M}{\frac{dU_{\mathrm{Skyrm}}}{dr}\, Y_{1M}}{0} \right|^2 \delta(\hbar\omega-\hbar\omega_{n,L=1})
 \nonumber \\
 &=\sum_{n}
 \left|
 \left\langle n,L=1 \left\lVert
 \frac{dU_{\mathrm{Skyrm}}}{dr}\,Y_{1}
 \right\rVert 0\right\rangle
 \right|^2
 \delta(\hbar\omega-\hbar\omega_{n,L=1}).
 \label{eq:strength-fcn-delV}
\end{align}
where in the second line we used the Wigner--Eckart theorem.
The resulting strength function is shown in Fig.~\ref{fig-strength_function}\,(c).

The coupling matrix element in Eq.~\eqref{eq:strength-fcn-delV} is extracted by fitting each peak of the strength function with a Lorentzian profile.
The value obtained for the lowest phonon state is listed in Table~\ref{table-strength}.
The extracted value 
\begin{equation}
 \left|
 \left\langle n,L=1 \left\lVert
 \frac{dU_{\mathrm{Skyrm}}}{dr}\,Y_{1}
 \right\rVert 0\right\rangle
 \right|^2
= 3.808 \times 10^{-3}\,\mathrm{MeV}^2\mathrm{fm}^{-2}
\end{equation}
reflects the suppression of the phonon amplitude shown in Fig.~\ref{fig-transition_density}\,(b).
This value can be compared with an estimate obtained by replacing the phonon amplitude with the unperturbed one given by Eq.~\eqref{eq:trans_analytic} and by the dotted curve in Fig.~\ref{fig-transition_density}\,(a).
This estimate yields $1.037 \times 10^{-1}\,\mathrm{MeV}^2\mathrm{fm}^{-2}$, which is listed in the third line of Table~\ref{table-strength}.

We thus find a strong suppression by approximately a factor of $30$.
Since the phonon amplitude is strongly suppressed around the surface and inside the cluster [see Fig.~\ref{fig-transition_density}\,(b)], the spatial overlap with the coupling field [Fig.~\ref{fig-transition_density}\,(c)] becomes much smaller than in the case of the unperturbed phonon amplitude.

These results provide the microscopic input for the matching condition~\eqref{Eq:matching condition2}, as discussed below.
In particular, the strong suppression of the coupling matrix element indicates that the interaction between the nuclear cluster and the superfluid phonon is much weaker than that expected from the unperturbed phonon profile.

\begin{table}
\centering
\caption{
 Properties of uniform neutron superfluid obtained from BCS calculations using the Skyrme energy density functional SLy4 and the density-dependent delta interaction for the pairing functional.
 The chemical potential $\mu_n$, neutron number density $\rho_n$, pairing gap $\Delta_n$, density response coefficient $\frac{\del\mu_n}{\del \rho_n}$ (inverse neutron susceptibility), and sound velocity $v_s$ (in units of the speed of light $c$) are listed.
 }
\label{table-uniform BCS}
\begin{tabular}{cc c cc} \\ \hline \hline
$\mu_n$(MeV) & $\rho_n$(fm${}^{-3}$) & $\Delta_n$(MeV) & $\frac{\del\mu_n}{\del \rho_n}$(MeVfm${}^{3}$) & $v_s/c$  \\ \hline
 0.5 & 1.81$\times 10^{-4}$ & 0.26  & 1.69$\times 10^{3}$  & 1.82$\times 10^{-2}$ \\
  1.0 & 5.54$\times 10^{-4}$ & 0.56  & 1.11$\times 10^{3}$  & 2.57$\times 10^{-2}$ \\
   2.0 & 1.70$\times 10^{-3}$ & 1.13  & 7.09$\times 10^{2}$  & 3.61$\times 10^{-2}$ \\
   4.0 & 5.47$\times 10^{-3}$ & 2.00  & 4.15$\times 10^{2}$  & 4.93$\times 10^{-2}$ \\
   8.0 & 2.09$\times 10^{-2}$ & 2.30  & 1.74$\times 10^{2}$  & 6.19$\times 10^{-2}$ \\
\hline \hline
\end{tabular}
\end{table}

\begin{table*}
\centering
\caption{
  Excitation energy $\hbar\omega$ of the lowest dipole superfluid phonon mode calculated with a spherical box of radius $R=50$ fm.
  The first row shows the result for the configuration with a nuclear cluster with $Z=28$ at the origin of the box, while the second row corresponds to the uniform neutron superfluid without the cluster.
  The second and third columns list the dipole matrix element $\left| \rbraketS{n,L=1}{rY_1}{0} \right|^2$ and the phonon coupling matrix element 
  $\left| \rbraketS{n,L=1}{\frac{dU_{\mathrm{Skyrm}}}{dr}Y_{1}}{0} \right|^2$, respectively.
  The third row shows the result obtained analyticaly 
  from the effective theory (Appendix A) for the configuration without a nuclear cluster. The diple and phonon coupling
  matrix elements are evaluated using the unperturbed phonon amplitude given in Eq.(\ref{eq:trans_analytic}).
  }
\label{table-strength}
\begin{tabular}{c ccc} \\ \hline \hline
    & \ $\hbar\omega$~(MeV) \ 
    & \ $\left| \rbraketS{n,L=1}{rY_1}{0} \right|^2$~(fm$^2$) \
    & \ $\left| \rbraketS{n,L=1}{\frac{dU_{Skyrm}}{dr}Y_{1}}{0} \right|^2$~(MeV$^2$fm$^{-2}$) \ \\ \hline
 with cluster (QRPA) & 0.941  &  $1.512 \times 10^5$ & $3.808 \times 10^{-3}$\\
 without cluster (QRPA) & 0.931 &  $1.476 \times 10^5$ & - \\
 without cluster (analytic) & 0.872 & $0.975 \times 10^5$ & $1.037 \times 10^{-1}$ \\
\hline \hline
\end{tabular}
\end{table*}

\subsection{Results}
\label{sec:results}

We are now in a position to determine the strength parameter $v_e$ of the effective potential $\hat{v}_{\mathrm{eff}}$.
The preceding subsections have provided the two ingredients needed for the matching: the matrix element on the effective-theory side has been expressed in terms of $v_e$, while the corresponding microscopic matrix element has been extracted from the linear-response calculation. 
For the size parameter $b$ of $\hat{v}_{\mathrm{eff}}$, we put $b=5.54$ fm, which is a surface radius of the nuclear cluster, evaluated by fitting
to the neutron density with a function of the Woods-Saxon shape\cite{Okihashi:2020kqg}.

If the microscopic and effective descriptions reproduced exactly the same phonon wave function, one could impose the matching condition Eq.~\eqref{Eq:matching condition2} directly.
However, as discussed in Sec.~\ref{sec:microscopic-side}, the QRPA calculation slightly overestimates the phonon amplitude compared with the analytic phonon solution, although the spatial profile of the transition density is reproduced reasonably well.
For this reason, we adopt a practical matching prescription in which the microscopic coupling matrix element is rescaled by the ratio of the dipole strengths as follows:
\begin{equation}
 \left|
 \left\langle n,L=1 \left\lVert
 \frac{dv_{\mathrm{eff}}}{dr}
 \right\rVert 0\right\rangle_{\mathrm{eff},S}
 \right|^2
 = 
 \left|
 \left\langle n,L=1 \left\lVert
 \frac{dU_{{\mathrm{Skyrme}}}}{dr}
 \right\rVert 0\right\rangle_{S}
 \right|^2
 \frac{\left| \rbraketpS{n,L=1}{rY_1}{0} \right|^2}
 { \left| \rbraketS{n,L=1}{rY_1}{0} \right|^2}
\end{equation}
The factor in the right-hand side compensates for the difference in the overall phonon amplitude between the microscopic and effective descriptions.

Using the numerical values listed in Table~\ref{table-strength}, we obtain $v_e= 7.47$ MeV.
Repeating the same analysis for different box sizes, $R=40$ and $60~\mathrm{fm}$, we find $v_e= 7.53$ and $7.98$ MeV, respectively.
Since these values are all close to one another, we adopt
\begin{equation}
v_e \approx 7.5~\mathrm{MeV}
\end{equation}
as a representative estimate.

It is also useful to compare the resulting effective interaction with the naive estimate based directly on the microscopic mean field.
For the effective potential, the zero-momentum component is
\begin{equation}
 \bar{v}_{\mathrm{eff}}(0)
= v_e b^3 \pi^{3/2} = 7.1 \times 10^3~\mathrm{MeV\,fm^3}.
\end{equation}
By contrast, if one uses the neutron Hartree--Fock potential itself in place of the effective interaction, the corresponding quantity is
\begin{equation}
\bar{v}_0(0)
=
-4.7 \times 10^4~\mathrm{MeV\,fm^3},
\end{equation}
whose magnitude is about seven times larger than $\bar{v}_{\mathrm{eff}}(0)$.
This indicates that the effective interaction between the nuclear cluster and the superfluid phonon may be substantially weaker than a naive estimate based on the Hartree--Fock potential.

\section{Discussion}
\label{sec:discussion}

\subsection{Dispersion relation and comparison with hydrodynamical approach}

The effective Hamiltonian~\eqref{Eq:Heff} contains a coupling between the superfluid phonon and the lattice vibration of the nuclear clusters.
As a consequence, the two modes are mixed to form two collective excitations.
The dispersion relation of the coupled modes is obtained as
\begin{equation}
\label{Eq:dispersion1}
 (\omega^2 -\omega_{\veck}^{l2})(\omega^2 - \omega_{\veck}^{s2}) = g_0^2 v_s^2 k^4 .
\end{equation}
This relation describes the mixing between the lattice phonon and the superfluid phonon.
Here, $\omega_{\veck}^{s}=v_s k$ and $\omega_{\veck}^{l}=v_l k$ denote the dispersion relations of the bare superfluid and lattice phonons, respectively.
The quantity $v_s$ is the velocity of the superfluid phonon in uniform neutron matter [recall Eq.~\eqref{eq:vs-superfluid-phonon}] while $v_l$ denotes the velocity of the lattice phonon.

In the present effective theory the coefficient of the $k^4$ term is determined by the interaction between the nuclear cluster and the surrounding neutrons.
Using the effective interaction $\bar v_{\mathrm{eff}}(0)$ introduced in Sec.~\ref{sec:long-wave-length-limit}, the coefficient can be written as
\begin{equation}
g_0^2 v_s^2=
\frac{\rho_{cl}\rho_n}{Mm}
|\bar v_{\mathrm{eff}}(0)|^2 ,
\end{equation}
where $\rho_{cl}$ denotes the cluster number density and $\rho_n$ is the neutron density outside the cluster in the surrounding superfluid.
The dispersion relation therefore shows that the mixing between the lattice vibration and the superfluid phonon is controlled by the effective interaction between the cluster and the surrounding neutrons.

The structure of the dispersion relation Eq.~\eqref{Eq:dispersion1} is similar to that obtained in hydrodynamical descriptions of the neutron-star crust~\cite{Pethick:2010zf,Chamel:2012ix,Chamel:2013cva,Kobyakov:2013eta}.
In the hydrodynamical formulation~\cite{Kobyakov:2013eta} the dispersion relation of the coupled modes takes the same form,
\begin{equation}
(\omega^2-\omega_{\veck}^{l2})(\omega^2-\omega_{\veck}^{s2})
=
v_{\mathrm{np}}^4 k^4 ,
\end{equation}
where the coupling constant $v_{\mathrm{np}}$ is expressed in terms of thermodynamic derivatives of the energy density as
\begin{equation}
v_{\mathrm{np}}^4
=
\frac{n_n^s}{m^2(n_n^n+n_p)}
(E_{nn}n_n^n + E_{np}n_p)^2 .
\end{equation}
Here $n_n^s$ and $n_n^n$ denote the number densities of superfluid and normal neutrons, respectively, while $n_p$ is the proton density.
The coefficients $E_{ij}$ are defined as
\begin{equation}
E_{nn}
=
\frac{\partial\mu_n}{\partial n_n}
=
\frac{\partial^2E}{\partial n_n^2},
\quad
E_{np}
=
\frac{\partial\mu_n}{\partial n_p}
=
\frac{\partial^2E}{\partial n_n\partial n_p}.
\end{equation}

In the hydrodynamical description the coupling term contains contributions from both the entrainment effect and the neutron--proton interaction.
In the present formulation the nuclear cluster is treated as a rigid impurity embedded in the neutron superfluid and the entrainment effect is not explicitly included. 
Neglecting the entrainment effect for comparison, the normal neutrons are
identified to those bound in clusters. As
a cluster contains $Z$ protons and $N$ neutrons bound inside the cluster,
the proton and neutron densities  may therefore be approximated as
$n_p = Z\rho_{cl}$ and $n_n^n = N\rho_{cl}$.
With these identifications, we find a correspondence
\begin{equation}
  |\bar v_{\mathrm{eff}}(0)| \leftrightarrow   N E_{nn}+Z E_{np},
\end{equation}
both of which characterizes the effective interaction between the cluster constituents and the surrounding neutrons.

Both formulations therefore describe the same physical coupling between the nuclear cluster and the neutron superfluid.
While the hydrodynamical approach expresses this interaction in terms of thermodynamic derivatives of the bulk energy density, the present effective theory provides a microscopic determination through the effective interaction $\bar v_{\mathrm{eff}}(0)$ obtained from the microscopic calculation.
The present framework thus provides a microscopic interpretation of the coupling parameter appearing in hydrodynamical descriptions of the neutron-star crust.

\subsection{Quantitative comparison}

We now examine the magnitude of the interaction strength $\bar v_{\mathrm{eff}}(0)$ (or equivalently the coupling parameter $g_0$) obtained in the present calculation and compare it with other estimates.

From the matching procedure described in Sec.~\ref{sec:results}, we obtained $\bar v_{\mathrm{eff}}(0)=7.1\times10^3\,\mathrm{MeV\,fm^3}$.
Using Eq.~\eqref{Eq:coupling constant}, the corresponding coupling parameter $g_0$ can be estimated by adopting the Wigner--Seitz radius $R_{WS}\approx28\,\mathrm{fm}$ from Ref.~\cite{Grill:2011dr} for a nuclear cluster with $Z=28$ and $A=88$. 
This yields
\begin{equation}
g_0/c = 4.3\times10^{-3}.
\end{equation}
This value is much smaller than the superfluid phonon velocity
$v_s/c = 4.93\times10^{-2}$ (Table \ref{table-uniform BCS}).
The relation $g_0\ll v_s$ indicates that the mixing between the lattice phonon and the superfluid phonon is weak.

It is instructive to compare this result with a naive estimate based directly on the neutron Hartree--Fock potential.
If the distortion of the phonon wave function is neglected and the interaction strength is approximated by the volume integral of the Skyrme--Hartree--Fock potential, one obtains $\bar{v}_{\mathrm{Skyrme}}(0)=-4.7 \times 10^4$ MeV fm$^3$. 
Our result is smaller by about a factor of seven compared with this crude estimate.

A similar comparison can be made with the hydrodynamical model of Ref.~\cite{Kobyakov:2013eta}.
From Table I of that work we find $E_{nn}=4.553\times10^3\,\mathrm{MeV\,fm^3}$ and $E_{np}=-1.853\times10^3\,\mathrm{MeV\,fm^3}$ for the neutron density outside the cluster $n^{\mathrm{out}}=5.777\times10^{-3}\,\mathrm{fm^{-3}}$.
This density is close to the density $\rho_n=5.47\times10^{-3}\,\mathrm{fm^{-3}}$ used in the present calculation. 
Using $Z=45.3$ and $N=109.3$ given in the same table, one obtains $NE_{nn} + ZE_{np}= -3.4 \times 10^4$ MeV fm$^3$.
This value is larger than our result
$\bar v_{\mathrm{eff}}(0)=7.1\times10^3\,\mathrm{MeV\,fm^3}$
by roughly a factor of five.

The reduction of the effective interaction strength compared with these estimates can be understood as a consequence of the strong distortion of the superfluid phonon in the vicinity of the nuclear cluster.
As shown in the microscopic calculation, the phonon amplitude is strongly suppressed inside and around the cluster.
Because the coupling matrix element is proportional to the overlap between the phonon wave function and the cluster potential, this suppression leads to a substantial reduction of the effective interaction.

For completeness we also compare our result with the estimate by Aguilera \textit{et al.}~\cite{Aguilera:2008ed}.
They approximated the ion (cluster)-neutron interaction as $v=2\pi \hbar^2 a_{nI}/m$ in terms of the scattering length $a_{nI}$.
Using their estimates $a_{nI}=10\,\mathrm{fm}$ or $\tilde a_{nI}=1.6\,\mathrm{fm}$ including medium effects, one finds $v=2.6\times10^3$ or $4.2\times10^2\,\mathrm{MeV\,fm^3}$.

These values are smaller than our result by a factor of a few to an order of magnitude.
However, the use of the scattering length may not be appropriate in the present context.
The scattering length characterizes the asymptotic behavior of the scattering wave function, whereas the cluster-neutron interaction relevant here occurs mainly in the vicinity of the nuclear surface.

\section{Conclusion} 
\label{sec:conclusion}

In this paper, we have investigated the coupling between the lattice vibration of nuclear clusters and the superfluid phonon in the inner crust of neutron stars.
Starting from a microscopic description based on the linear-response quasiparticle random-phase approximation (QRPA) within a nuclear energy density functional, we derived the interaction between the nuclear cluster and the surrounding neutron superfluid.
We then formulated an effective description of the coupled lattice and superfluid phonon modes in the
long-wavelength limit in terms of an effective Hamiltonian
including an effective coupling between lattice and superfluid phonons.
By comparing the microscopic response with the 
long-wavelength effective  description, we identified the matching condition that determines the effective coupling constant from the microscopic phonon coupling matrix element.
This procedure allows the coupling parameter appearing in the effective Hamiltonian to be determined quantitatively from the microscopic calculation.

Our microscopic calculation shows that the effective coupling strength is significantly reduced compared with naive estimates.
This reduction originates from the strong suppression of the superfluid phonon amplitude inside and in the vicinity of the nuclear cluster.
Because the coupling matrix element is proportional to the overlap between the phonon wave function and the cluster potential, the distortion of the phonon wave function leads to a substantial decrease of the effective coupling strength.

The resulting coupling strength is smaller than the estimate based on the hydrodynamical model of Ref.~\cite{Kobyakov:2013eta} by roughly a factor of five when the entrainment effect is neglected.
The present analysis therefore provides a microscopic interpretation of the coupling parameter appearing in hydrodynamical descriptions of the neutron-star crust.

In the present framework, the entrainment effect has not been included explicitly.
Previous studies~\cite{Cirigliano:2011tj,PageReddy2012,Chamel:2012ix,Chamel:2013cva} have suggested that the entrainment effect may enhance the coupling between the lattice and the superfluid component.
It will therefore be important to investigate the entrainment contribution within a microscopic framework, for example by extending the present DFT-based linear-response approach.
Such an analysis would allow a more quantitative connection between microscopic nuclear models and hydrodynamical descriptions of neutron-star crust dynamics.

\appendix

\acknowledgments

M.M. is supported by the Japan Society for the Promotion of Science (JSPS) KAKENHI Grants No. 20K03945 and No. 24K07014.
M.H. is supported by the Japan Society for the Promotion of Science (JSPS) KAKENHI Grants No. 23K25870, No. 25K01002, No. 25K07316, and No. 26H01407.
This work was partially supported by the COREnet project of RCNP at The University of Osaka, RIKEN iTHEMS, and Niigata University Quantum Research Center (NU-Q).

\section{Effective Hamiltonian for superfluid phonon}
\label{sec:note}

The effective Hamiltonian for low-frequency excitations of a superfluid system consisting of particles with mass $m$ is given by~\cite{Khalatnikov,Lifshitz-Statphys2}
\begin{equation}
\hat{H}_{\mathrm{eff}}
=
\int d\vecr
\left[
\frac{\hbar^2\rho}{2m}
\big(\nabla\hat{\varphi}(\vecr)\big)^2
+
\frac{1}{2}
\frac{\partial\mu}{\partial\rho}
\big(\delta\hat{\rho}(\vecr)\big)^2
\right].
\end{equation}
Here, $\hat{\varphi}(\vecr)$ represents the phase of the superfluid order parameter, while $\delta\hat{\rho}(\vecr)$ denotes the fluctuation of the number density. 
These operators satisfy the canonical commutation relation $[\hat{\varphi}(\vecr), \delta\hat{\rho}(\vecr')] = i \delta(\vecr-\vecr')$. 
The quantity $\rho$ denotes the number density in the ground state, and
\begin{equation}
 \frac{\partial\mu}{\partial\rho}
 = \frac{\partial^2E}{\partial\rho^2}
\end{equation}
is the second derivative of the ground-state energy density $E$ with respect to the number density. 
This quantity can be identified as the inverse number susceptibility (compressibility).

Introducing annihilation and creation operators for the phonon eigenmodes,
$\hat{b}_i$ and $\hat{b}_i^{\dagger}$, the Hamiltonian, the phase operator, and the density fluctuation operator can be written as
\begin{align}
 \hat{H}_{\mathrm{eff}} 
 &= \sum_i  \hbar\omega_i \left(\hat{b}_i^{\dagger}\hat{b}_i + \frac{1}{2}\right), 
 \\
 \hat{\varphi}(\vecr)
 &= \sum_i \phi_i(\vecr)\sqrt{  \frac{\frac{\del\mu}{\del\rho}}{2\hbar\omega_i}  } \left(\hat{b}_i+\hat{b}_{-i}^{\dagger}\right),
 \\
 \delta\hat{\rho}(\vecr) 
 &= \sum_i  \phi_{-i}(\vecr)\sqrt{\frac{\hbar\omega_i}{2\frac{\del\mu}{\del\rho}}} i \left(\hat{b}^\dagger_i-\hat{b}_{-i}\right),
\end{align}
where the subscript $-i$ denotes the time-reversed state of $i$, namely
\begin{equation}
\phi_{-i}(\vecr)=\phi_i^{*}(\vecr).
\end{equation}

The phonon amplitude $\phi_i(\vecr)$ of the eigenmodes obeys the wave equation
\begin{equation}
 \omega_i^2 \phi_i(\vecr)
 = v^2 \nabla^2 \phi_i(\vecr)
\end{equation}
subject to appropriate boundary conditions, where the sound velocity is
\begin{equation}
 v=\sqrt{\frac{\rho}{m}\frac{\partial\mu}{\partial\rho}}.
\end{equation}

For a large rectangular box $V$ with periodic boundary conditions, the eigenmodes are labeled by the momentum $\veck$, and the phonon wave function is
\begin{equation}
 \phi_{\veck}(\vecr)
 = \frac{1}{\sqrt{V}} e^{i\veck\cdot\vecr},
\end{equation}
with the dispersion relation
\begin{equation}
\omega_{\veck}=vk .
\end{equation}

For phonons confined in a spherical box $S$ of radius $R$, the eigenmodes are specified by radial and angular quantum numbers,
\begin{equation}
\phi_{nLM}(\vecr)
=
\frac{1}{R^{3/2}}
c_{nL}
j_L(k_n r)
Y_{LM}(\hat{\vecr}),
\end{equation}
where
\begin{equation}
c_{nL}
=
(x_{nL})^{3/2}
\left(
\int_0^{x_{nL}}
x^2
j_L^2(x)
dx
\right)^{-1/2}
\end{equation}
is a normalization constant.
The frequency is given by
\begin{equation}
\omega_{k_n}=vk_n ,
\end{equation}
with $k_n=x_{nL}/R$, where $x_{nl}$ denotes the $n$-th zero of the spherical Bessel function $j_L(x)$.

If the particles are fermions and their interaction is neglected,
the energy density can be approximated by that of a free Fermi gas.
In this case, one obtains
\begin{equation}
\frac{\partial\mu}{\partial\rho}
=
\frac{\hbar^2\pi^2}{m k_F},
\end{equation}
and the sound velocity becomes
\begin{equation}
v=\frac{1}{\sqrt{3}}v_F.
\end{equation}
Note that the coefficient $\partial\mu/\partial\rho$ increases as the density decreases.

\bibliographystyle{utphys}
\bibliography{refs}
\end{document}